\documentclass[aps,prb,twocolumn,floats,amsmath,amssymb,superscriptaddress,noeprint,longbibliography]{revtex4-2}

\usepackage{here}
\usepackage{subcaption}
\usepackage{tabularx}
\usepackage{graphicx}
\usepackage{bm}
\usepackage{dcolumn}
\usepackage{amsthm}
\usepackage{tikz}
\usepackage{siunitx}
\DeclareSIUnit\angstrom{\text {Å}}
\usepackage{hyperref}
\usepackage{booktabs}
\usepackage{threeparttable}  
\usepackage{multirow}
\usepackage{dcolumn}
\usepackage{blindtext}
\hypersetup{
  colorlinks,
  citecolor=blue,
  linkcolor=blue,
  urlcolor=blue}

\newcommand{\blue}[1]{{\color{black}#1}} 

\newcommand{\eg}{{\it e.g.}, }
\newcommand{\ie}{{\it i.e.}, }

\begin{document}

\title{Structure and thermodynamic stability of $\beta$-Ga$_2$O$_3$ surfaces}

\author{Konstantin Lion}
\thanks{Current address: Molecular Simulations from First Principles e.V., Berlin, Germany}
\affiliation{Institut f\"ur Physik and CSMB, Humboldt-Universit\"at zu Berlin, Berlin, Germany}
\affiliation{Fritz-Haber-Institut der Max-Planck-Gesellschaft, Berlin, Germany}
\author{Claudia Draxl}
\affiliation{Institut f\"ur Physik and CSMB, Humboldt-Universit\"at zu Berlin, Berlin, Germany}
\date{\today}

\begin{abstract}
We present a comprehensive first-principles investigation of all symmetrically inequivalent low-index surfaces of $\beta$-Ga$_2$O$_3$, examining their structural properties and thermodynamic stability across experimentally relevant growth conditions. Using density-functional theory with both \blue{the} semi-local \blue{functional PBEsol and a modified PBE0 hybrid functional with \SI{26}{\percent} exact exchange, denoted PBE0(0.26)}, we calculate surface free energies for the (010), (100), (001), ($\bar{2}01$), (110), (111), and ($11\bar{1}$) orientations, including the effects of harmonic vibrational contributions and varying oxygen chemical potentials. We demonstrate that the energetic ordering remains consistent across computational approaches and that the vibrational contributions remain below \SI{0.2}{\joule\per\meter^2} up to temperatures of \SI{1000}{\kelvin}. A coordination-based model that correlates surface stability with the density of under-coordinated atoms reveals that under-coordinated oxygen atoms and tetrahedral Ga sites substantially destabilize surfaces, while exposed under-coordinated octahedral Ga atoms serve as indicators of surface stability. Our thermodynamic analysis shows that stoichiometric terminations dominate over nearly the entire range of chemical potentials relevant for $\beta$-Ga$_2$O$_3$ stability, while non-stoichiometric terminations emerge only under extreme reducing or oxidizing conditions. Notably, we predict the formation of stable Ga-rich terminations resembling Ga adlayers for the (100) and ($\bar{2}01$) surfaces under highly reducing conditions.
\end{abstract}
\maketitle

\section{Introduction}

\begin{figure} \centering
    \includegraphics[width=0.35\textwidth]{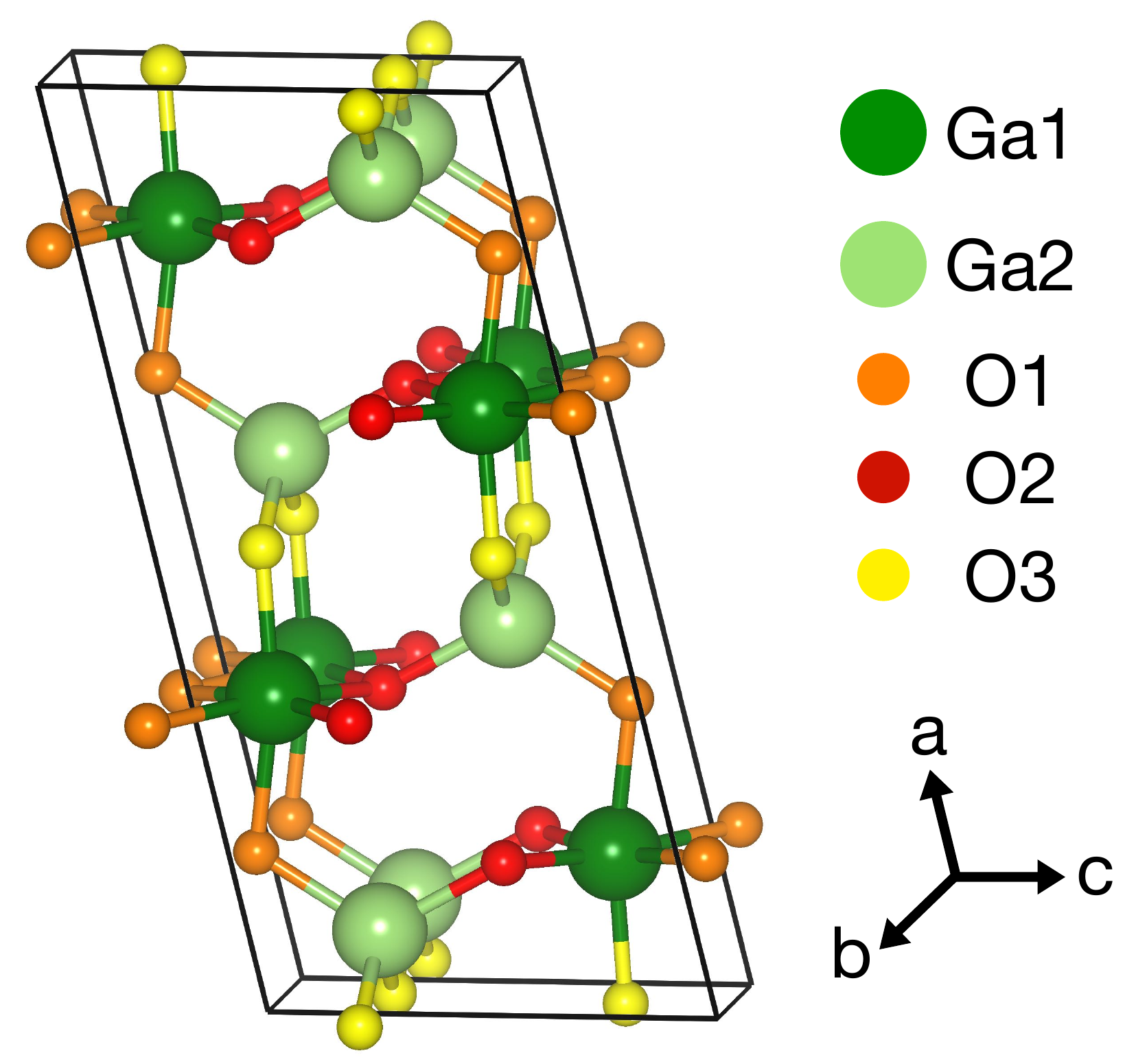}
    \caption{Conventional bulk unit cell of $\beta$-Ga$_2$O$_3$ consisting of 2 inequivalent gallium atoms and 3 inequivalent oxygen atoms. The octahedrally (tetrahedrally) coordinated gallium atoms, Ga1 (Ga2), are dark (light) green. The inequivalent oxygen atoms are shown in orange (O1), red (O2), and yellow (O3).\label{fig:bulk}}
\end{figure} 

Gallium oxide, Ga$_2$O$_3$, has emerged as a promising ultrawide-bandgap semiconductor for high-power electronics and gas-sensing applications due to its exceptional electrical and optical properties~\cite{green2022,higashiwaki2018,higashiwaki2020,pearton2018}. Among the five known polymorphs, the monoclinic $\beta$-phase is the thermodynamically most stable under ambient conditions and can be grown as large single crystals from the melt~\cite{galazka2017,galazka2018}. The successful implementation of $\beta$-Ga$_2$O$_3$ in electronic devices critically depends on controlled crystal growth and surface engineering. In general, surface properties directly influence crystal morphology during growth, chemical reactivity, and electronic characteristics at metal-semiconductor and oxide-semiconductor interfaces, all essential factors for device fabrication and performance. 

The monoclinic crystal structure of $\beta$-Ga$_2$O$_3$ (space group C$_{2}m$), shown in Fig.~\ref{fig:bulk}, contains two inequivalent gallium sites (tetrahedrally coordinated Ga2 and octahedrally coordinated Ga1) and three inequivalent oxygen atoms, giving rise to numerous possible surface terminations with distinct atomic arrangements and properties. Density-functional theory~(DFT) calculations have consistently identified the (100) surface as the most stable low-index orientation~\cite{bermudez2006,mu2020}, a result that holds across different exchange-correlation functionals and has been extended to include higher-index orientations~\cite{hinuma2019a,hinuma2020a}. Beyond stoichiometric terminations, individual studies have examined hydrogen adsorption and oxygen vacancies on (100)~\cite{gonzalez2005}, surface reconstructions on (010) and (110)~\cite{wang2023}, and water activation on ($\bar{2}$01)~\cite{anvari2018}, while comparisons across Ga$_2$O$_3$ polymorphs have been performed under epitaxial strain~\cite{bertoni2024}. \blue{Recently, we have explored reconstructions of the (001) surface beyond bulk truncation by replica-exchange grand-canonical molecular dynamics and \textit{ab initio} atomistic thermodynamics, revealing a $1\times2$ reconstruction of paired GaO$_4$ tetrahedra, consistent with high-angle annular dark-field scanning transmission electron microscopy~\cite{lion2026}.} A unified picture covering all symmetrically inequivalent low-index orientations is, however, still missing.

Experimentally, the link between surface energetics and crystal growth is well established. Metal-organic vapor phase epitaxy (MOVPE) on (100) substrates has demonstrated step-flow growth mechanisms~\cite{anooz2020,anooz2020a,schewski2019a} and systematic evolution of surface morphology with growth parameters~\cite{cheng2018}, consistent with the predicted low surface energy of this orientation. \blue{Surface x-ray diffraction of as-cleaved (100) single crystals confirms a single, bulk-like termination with atomic displacements of at most \SI{0.1}{\angstrom}~\cite{kao2026}.} Chou \textit{et al.}~\cite{chou2021,chou2023} showed that maintaining step-flow growth requires specific oxygen-to-gallium ratios, suggesting the formation of Ga-rich surface phases under reducing conditions, while growth on (010) substrates reveals distinct growth windows~\cite{bhattacharyya2020}. These observations highlight the need to understand surface energetics across the full range of thermodynamic conditions relevant to growth.

In this work, we present a comprehensive first-principles investigation of all symmetrically inequivalent low-index surfaces of $\beta$-Ga$_2$O$_3$, namely (010), (100), (001), ($\bar{2}$01), (110), (111), and (11$\bar{1}$). \blue{Although several of these orientations have been addressed in previous computational studies~\cite{bermudez2006,mu2020,hinuma2019a,hinuma2020a}, the available results were obtained with different DFT codes and exchange-correlation functionals and are, with few exceptions, restricted to stoichiometric terminations. Consequently, neither the dependence of the stable terminations on the growth conditions nor the sensitivity of the energetic ordering to the level of theory and to vibrational contributions is established. The aim of this work is to close this gap by treating all low-index orientations and all their bulk-truncated terminations within one consistent all-electron framework.} We employ both semi-local and hybrid-functional levels of DFT to compare their effect on surface stability, while also assessing the impact of vibrational contributions on thermodynamic stability. By examining the relative stability of these surfaces across a wide range of oxygen chemical potentials relevant to $\beta$-Ga$_2$O$_3$ growth and processing, we provide a comprehensive picture of the material's surface energetics. In addition, we develop a coordination-based model that correlates atomic structure with surface stability, which deepens our understanding of growth behavior and guides substrate selection for device applications.

\section{Theoretical background}
The surface free energy ($\gamma$) of a slab structure, which is influenced by temperature ($T$) and the partial pressures ($p_{i}$) of its constituent elements, is determined as~\cite{qian1988,rogal2004,reuter2003}
\begin{align}
    \gamma\left(T,\{p_{i}\}\right) = \frac{1}{2A} \left[G_{\text{slab}}(T,\{p_{i}\}) - \sum_i N_i \mu_i\left(T,p_{i}\right)\right] \, , \label{eq:basic_gamma}
    \end{align}
where $G_{\text{slab}}$ is the Gibbs free energy of the slab, $N_i$ the number of atoms of each element in the slab, and $\mu_i$ the respective chemical potential. The factor 2 in the denominator indicates that we are simulating symmetric slabs with equivalent surfaces at its top and bottom. The thermodynamic stability of surfaces, which exchange atoms with an oxygen reservoir, is determined by their surface free energies as defined in Eq.~\ref{eq:basic_gamma}. If we consider the surface to be in equilibrium with the underlying bulk and the gas phase~\cite{qian1988,reuter2001,rogal2007a}, then the chemical potentials of oxygen and gallium are related by
\begin{align}
    g_{\text{bulk}} = 2\,\mu_{\text{Ga}} + 3\,\mu_{\text{O}} \, , \label{eq:bulk_stability}
\end{align}
with $g_{\text{bulk}}$ being the Gibbs free energy of a bulk Ga$_2$O$_3$ formula unit. We can then rewrite Eq.~\ref{eq:basic_gamma} as:
\begin{equation} 
\begin{split}
    \gamma (T,p_{\text{O}}) = & \frac{1}{2A} \bigg[ G_{\text{slab}} -\left( N_{\text{O}}-\frac{3}{2}\, N_{\text{Ga}}\right)\,\mu_{\text{O}}( T,p_{\text{O}} )\\
    & -\frac{1}{2} N_{\text{Ga}}\,g_{\text{bulk}} \bigg] \, . 
\end{split}
\label{eq:delta_gamma_2}
\end{equation}

The Gibbs free energies are approximated by the total energies obtained by DFT, $G \approx E^{\text{DFT}}$, a common approach in literature~\cite{reuter2001,reuter2003}, which is justified by the inherent error cancellation of slab and bulk energies in Eq.~\ref{eq:delta_gamma_2}. We will, however, explicitly show for the stoichiometric terminations that including the vibrational free energy $F_{\text{vib}}$ can affect the relative stability between surfaces by the order of several \si{\milli\joule\per\meter^2}. When vibrational contributions are considered as part of the Gibbs free energy, we obtain an additional term contributing to the surface free energy:
\begin{equation} 
\begin{split}
    \gamma_{\text{vib}} & (T) = \frac{1}{2A} \bigg[ F^{\text{vib}}_{\text{slab}}(T) -\frac{1}{2} N_{\text{Ga}}\,f^{\text{vib}}_{\text{bulk}}( T) \bigg] \, .
\end{split}
\label{eq:delta_gamma_vib}
\end{equation}
The chemical potential of oxygen is referenced to its standard state, \ie molecular O$_2$ including the experimental zero-point energy~\cite{irikura2007}:
\begin{align}
    \mu_{\text{O}}(T,p_{\text{O}})&=\Delta\mu_{\text{O}}(T,p) + 1/2 E_{\text{O}_2} .
\end{align}
The binding energy of oxygen is known to be difficult to approximate with (semi)local DFT due to a significant self-interaction error~\cite{zhang1998, perdew1996, ernzerhof1997}. In this work, we obtain \SI{-6.76}{\electronvolt} using PBEsol and \SI{-5.24}{\electronvolt} using PBE0 with a mixing factor $\alpha=0.26$, termed PBE0(0.26). Additional details on the choice of these functionals are given in Section~\ref{sec:comp_details}. To reduce the uncertainty when comparing different functionals, the experimental value for the binding energy $E_{\text{b}}^{\text{exp}}=\SI{-5.22}{\electronvolt}$~\cite{feller1999} is used. The total energy of the molecule is then given by $E_{\text{O}_2}=2 E_{\text{O}} + E_{\text{b}}^{\text{exp}}$, where the free energy $E_{\text{O}}$ of the  oxygen atom is evaluated at the respective exchange-correlation level.  

We convert $\Delta\mu(T,p)$ into pressure and temperature conditions using thermodynamical tables~\cite{stull1971}. The bounds for the chemical potentials can be deduced from the condition of the thermodynamic stability of bulk Ga$_2$O$_3$. Below the O-poor (or Ga-rich) limit, the oxide will decompose into Ga metal and oxygen, while in the O-rich (or Ga-poor) limit, oxygen will condense on the surface. Reasonable bounds for the chemical potentials are thus given as~\cite{reuter2001}
\begin{align}
    \frac{1}{3}H_f(T=\SI{0}{\kelvin},p=0)&< \Delta\mu_{\text{O}} < 0 \, ,
\end{align}
where $H_f(T=\SI{0}{\kelvin},p=0)$ is the enthalpy of formation for $\beta$-Ga$_2$O$_3$, for which we obtain \SI{-9.64}{\electronvolt} using PBEsol and \SI{-9.98}{\electronvolt} using PBE0(0.26). \blue{Note that $H_f$ is evaluated consistently at each level of theory, \ie using the DFT total energy of the O$_2$ molecule without the experimental correction of the binding energy introduced above.} This underestimates the experimental value of \SI{-11.29}{\electronvolt}~\cite{lideed.} at ambient conditions\blue{. For PBEsol, this can partly be attributed to the overestimation of the oxygen binding energy mentioned above. The remaining discrepancy, which is also present for PBE0(0.26), may stem from the description of the metallic Ga reference and of the oxide itself, and from the fact that the experimental value is obtained at \SI{298}{\kelvin} and \SI{1}{\bar}, whereas $H_f(T=\SI{0}{\kelvin},p=0)$ is calculated from DFT total energies.} The limits for the chemical potentials are thus~-3.21 [-3.33]\,$~\si{\electronvolt}<\Delta\mu_{\text{O}}<\SI{0}{\electronvolt}$ for PBEsol~[PBE0(0.26)]. We note that these limits are not strict extrema and may even be exceeded in experiments via different oxygen sources, such as ozone or oxygen plasma. When presenting surface free energies, we will thus include a range slightly outside these limits to better compare to experimental values for the Gibbs free energy of formation.

\section{Computational details\label{sec:comp_details}}
All DFT calculations are performed using atomic-centered basis functions as implemented in the all-electron full-potential code FHI-aims~\cite{blum2009,abbott2026}. Exchange-correlation effects are treated (i) in the generalized gradient approximation by the revised PBE functional for solids and surfaces (PBEsol)~\cite{perdew2008}, which has shown high accuracy in determining the lattice parameters and elastic properties of Ga$_2$O$_3$~\cite{wouters2020,lion2022}
and (ii) with the hybrid functional PBE0~\cite{adamo1999,ernzerhof1999} including \SI{26}{\percent} exact exchange~\cite{deak2019} [denoted as PBE0(0.26)], which matches the experimental gap of \SI{4.9}{\electronvolt}~\cite{orita2000}. The Brillouin zone of the bulk structure is sampled with a $4\times 12 \times 8$ k-grid and scaled accordingly for the surface structures. The calculated bulk lattice parameters using PBEsol are $a=\SI{12.28}{\angstrom}$, $b=\SI{3.05}{\angstrom}$, $c=\SI{5.81}{\angstrom}$, and $\beta=\SI{103.72}{\degree}$, which agree very well with the experimental values~\cite{ahman1996} of $a=\SI{12.21}{\angstrom}$, $b=\SI{3.04}{\angstrom}$, $c=\SI{5.80}{\angstrom}$, $\beta=\SI{103.83}{\degree}$. The surface structures are created using pymatgen~\cite{ong2013,sun2013} and are modeled as slabs in the supercell approach, including at least \SI{50}{\angstrom} of vacuum. The resulting slabs are symmetric due to the inversion symmetry of the bulk structure, allowing us to extract the properties of the single surface. All atoms in the slab are relaxed with the final forces being smaller than \SI{E-2}{\electronvolt\per\angstrom}. In the PBEsol calculations, the basis set and numerical grids are defined by \textit{tight} species defaults for both the gallium and oxygen atoms, while for the PBE0(0.26) calculations, we employ \textit{intermediate} defaults. We verified the stability of all investigated terminations in larger surface supercells. A systematic exploration of further reconstructions and adsorbate-covered terminations is beyond the scope of this work.

The vibrational free energies for selected structures are calculated in the harmonic approximation using the Python package phonopy~\cite{togo2023a,togo2023b}. In the phonon calculations, all slabs are modeled in supercells with lateral dimensions of at least $\SI{12}{\angstrom} \times \SI{12}{\angstrom}$, as previously used in defect calculations of $\beta$-Ga$_2$O$_3$~\cite{deak2017}. The corresponding energies are obtained using the functional PBEsol and \textit{tight} species defaults.

The input and output files of all calculations, which enable reproducing the results, are openly available at NOMAD~\cite{draxl2018, draxl2019,scheidgen2023} under the DOI \url{http://doi.org/10.17172/NOMAD/2026.04.15-1}.

\section{Results}

\begin{figure} \centering
\begin{subfigure}[b]{0.47\textwidth}
   \includegraphics[width=\textwidth]{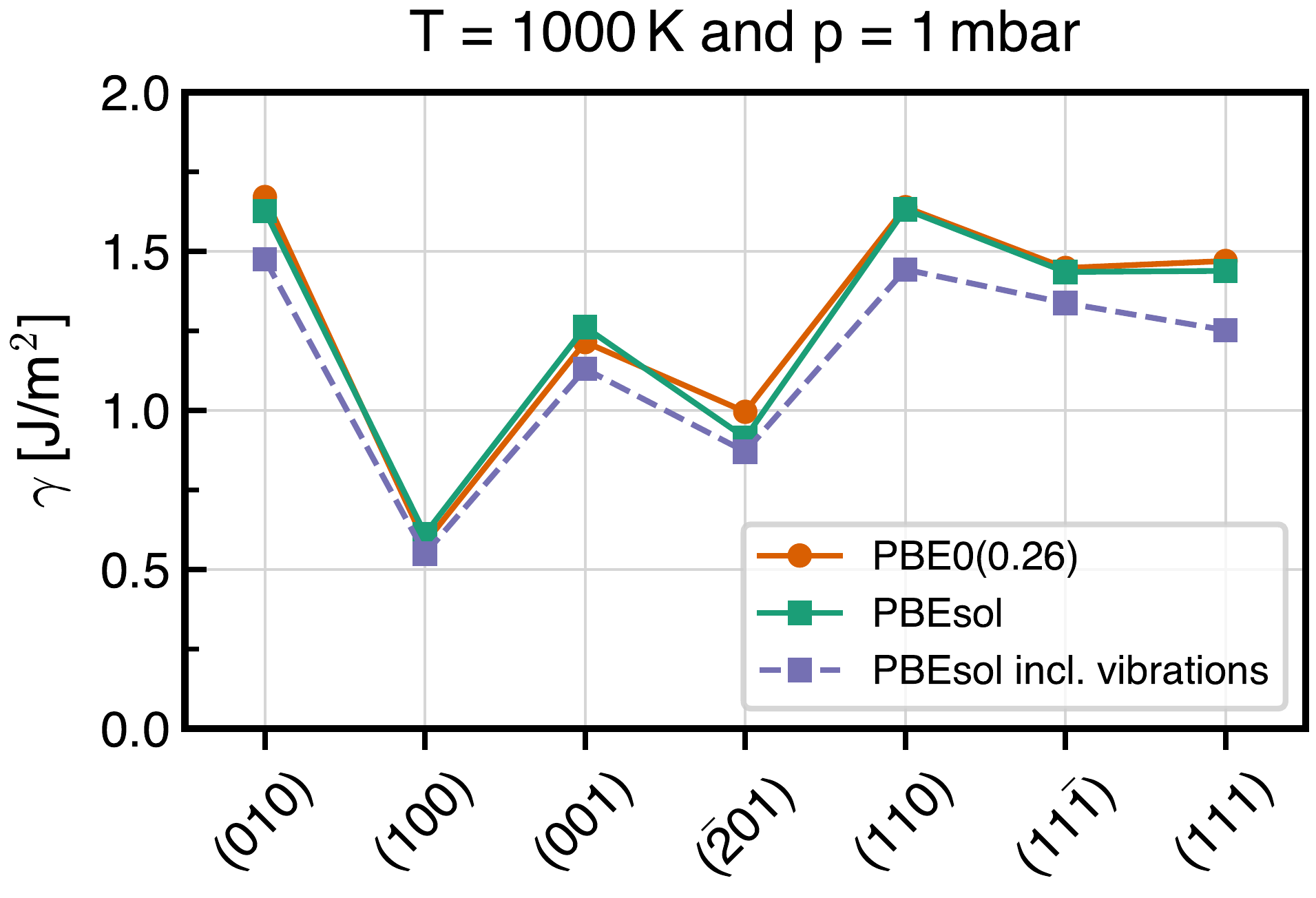}
   \label{fig:first_figure}
\end{subfigure}

\begin{subfigure}[b]{0.48\textwidth}
   \includegraphics[width=\textwidth]{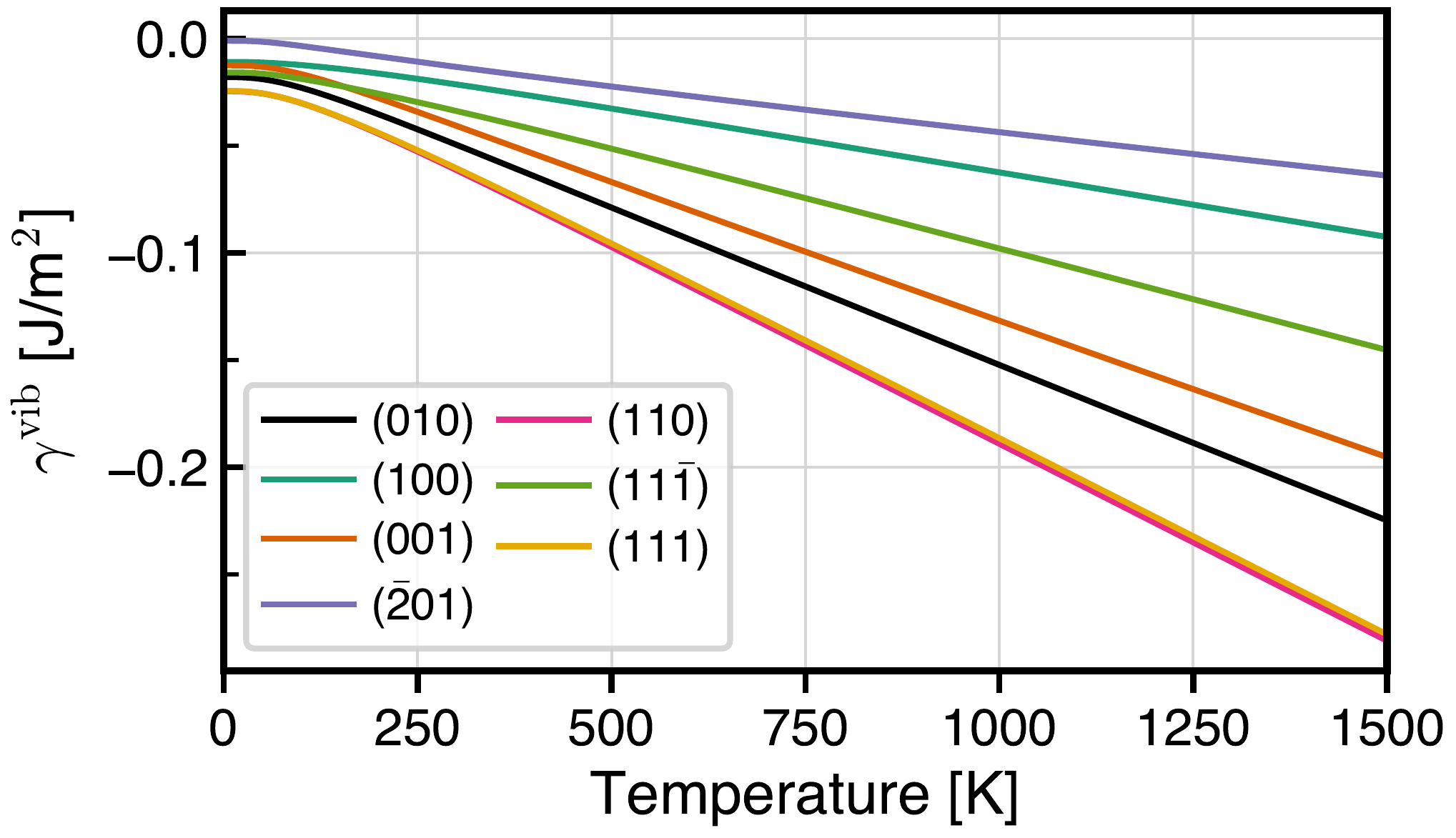}
   \label{fig:second_figure}
\end{subfigure}
\caption{Top: Surface free energies of the most stable stoichiometric terminations of $\beta$-Ga$_2$O$_3$ obtained with PBEsol, excluding (green symbols) and including (purple symbols) harmonic vibrational contributions, as well as PBE0(0.26) (orange symbols). The lines are guides to the eye. Bottom: Vibrational contributions to the surface free energies in the harmonic approximation for the same terminations as a function of temperature.
\label{fig:all_stoichiometric_surfaces_gamma}}
\end{figure} 


\subsection{Stability of stoichiometric terminations\label{sec:stoichiometric}}

We begin our analysis with the stoichiometric terminations of the low-index surfaces. Stoichiometric terminations maintain the gallium to oxygen ratio (2:3) of the bulk crystal. All investigated surface directions, except for (010), possess two distinct stoichiometric terminations. The (010) surface represents a special case with only one stoichiometric termination, because this direction is parallel to the C$_{2}m$ symmetry axis of the bulk crystal. \blue{All stoichiometric slabs considered in this section are centrosymmetric and therefore carry no net dipole moment. More fundamentally, in the classification of Tasker~\cite{tasker1979}, the stoichiometric bulk-truncated surfaces of all low-index orientations are of type 2, \ie they consist of charged atomic layers stacked such that the repeat unit has no net dipole, with the exception of (010), which is of type 1 with charge-neutral layers~\cite{hinuma2020a}. None of these stoichiometric surfaces is therefore polar in the sense of a Tasker type-3 surface, and no polarity compensation mechanism is required for their stability. In contrast, the non-stoichiometric (O-rich and Ga-rich) bulk-truncated terminations discussed in Section~\ref{sec:non-stoichiometric} can carry a formal surface charge, which contributes to their lower stability.}

The expression for the surface energy for a stoichiometric surface can be simplified by inserting the bulk gallium to oxygen ratio into Eq.~\ref{eq:delta_gamma_2}:
\begin{align}
    \gamma (T,p_{\text{O}}) = & \frac{1}{2A} \bigg[ G_{\text{slab}} -\frac{1}{2} N_{\text{Ga}}\,g_{\text{bulk}} \bigg] \, . 
    \label{eq:delta_gamma_stoichio}
\end{align}

Figure~\ref{fig:all_stoichiometric_surfaces_gamma} (top panel) shows the calculated surface free energies for the most stable stoichiometric terminations of each surface orientation at a typical growth temperature of \SI{1000}{\kelvin}~\cite{schewski2019a}. Results are shown for calculations performed with PBEsol and PBE0(0.26), as well as PBEsol including vibrational contributions in the harmonic approximation. The surface free energies range from approximately \SI{0.6}{\joule\per\meter^2} for the (100) surface to approximately \SI{1.7}{\joule\per\meter^2} for the (111) and (010) surfaces. A comparison with reported literature values is presented in Table~\ref{tab:sur_energies}. The energetic ordering of the surfaces in this work, (100) $<$ ($\bar{2}$01) $<$ (001) $<$ (111) $\approx$ (11$\bar{1}$) $<$ (110) $\approx$ (010), is consistent across all references. The absolute values show slight differences, which we attribute to the varying computational approaches employed, such as all-electron versus pseudopotential methods, and the use of different exchange-correlation functionals (GGA versus hybrid functionals).

\begin{table}[h]
\centering
\begin{threeparttable}
\caption{Relaxed surface energies in \si{\joule\per\meter^2} of the most stable stoichiometric terminations of $\beta$-Ga$_2$O$_3$ compared to literature values where available. \label{tab:sur_energies}}
\label{tab:surface_energies}
\begin{tabular}{ccccccc}
\hline
\hline
Surface & PBEsol & PBE0(0.26) & Ref.~\cite{mu2020}\tnote{a} & Ref.~\cite{bermudez2006}\tnote{b} & Ref.~\cite{hinuma2019a}\tnote{c} & Ref.~\cite{hinuma2020a}\tnote{c}\\
\hline
(100) & 0.61 & 0.58 & 0.34 & 0.68 & 0.38 & 0.47 \\
(010) & 1.63 & 1.67 & 1.67 & 2.03 & 1.49 & 1.60 \\
(001) & 1.26 & 1.22 & 1.17 & 1.40 & --- & 1.23 \\
($\overline{2}$01) & 0.96 & 1.00 & --- & --- & 0.75 & 0.77 \\
(110) & 1.63 & 1.64 & --- & --- & --- & 1.58 \\
(111) & 1.42 & 1.47 & --- & --- & --- & 1.26 \\
(11$\overline{1}$) & 1.43 & 1.45 & --- & --- & --- & 1.31 \\
\hline
\hline
\end{tabular}
\begin{tablenotes}
\item[a] DFT-HSE06 with a mixing parameter of 0.32
\item[b] DFT-B3LYP
\item[c] DFT-PBEsol
\end{tablenotes}
\end{threeparttable}
\end{table}

These results are applicable across growth conditions, since the surface free energy (excluding vibrational contributions and configurational entropy) for stoichiometric terminations is independent of the oxygen chemical potential (cf. Eq.~\ref{eq:bulk_stability}). The comparison between PBEsol and PBE0(0.26) calculations reveals only minor energetic differences. This suggests that while hybrid functionals are essential for the accurate description of the electronic structure in these systems (cf.~Appendix~\ref{sec:appendix}), PBEsol adequately captures the energetic ordering of these surfaces. The good performance of PBEsol can be attributed to the fact that we evaluate energetic differences between similar systems (slab and bulk), which benefits from significant error cancellation.

The inclusion of vibrational contributions to the surface free energy introduces small but noticeable modifications to the relative stability of certain surfaces. Note that the vibrational contributions, as calculated in Eq.~\ref{eq:delta_gamma_vib}, are still temperature dependent. As shown in Fig.~\ref{fig:all_stoichiometric_surfaces_gamma} (bottom panel), the vibrational contributions are uniformly negative and become more significant with increasing temperature, reaching values between \SIrange{-0.05}{-0.2}{\joule\per\meter^2} at \SI{1000}{\kelvin}. These contributions arise from the modified vibrational density of states at the surface compared to the bulk, primarily due to the reduction in coordination and subsequent changes in bond strengths. Notably, the (110) and (111) surfaces show the largest vibrational contributions, while the ($\bar{2}$01) surface exhibits the smallest effect. The relative stability, however, is not affected when including vibrational effects. Given this small impact, we do not include them in the subsequent analysis. For studies focusing on closely competing surface terminations with energy differences smaller than \SI{0.1}{\joule\per\meter^2}, inclusion of vibrational effects might be necessary for accurate predictions of the relative stability.

To gain physical insight into the factors affecting surface stability in this material, we develop a coordination-based model similar to the one used for LiFePO$_4$~\cite{wang2007}. It correlates the PBEsol surface energies shown in Fig.~\ref{fig:all_stoichiometric_surfaces_gamma} with the density of under-coordinated atoms at the termination layer of each surface. For the stoichiometric surface terminations studied, we identify the atoms at the outermost surface layer, which have one or more dangling bonds compared to their bulk coordination. We find that in the fully relaxed surface terminations, only specific under-coordinated environments occur: 3-fold tetrahedrally coordinated Ga2 (reduced from 4-fold in bulk), 4-fold and 5-fold octahedrally coordinated Ga1 (reduced from 6-fold in bulk), and 2-fold coordinated O atoms (reduced from 3-fold in bulk). The following model expresses the surface energy $\gamma$ as a linear combination of the corresponding contributions,
\begin{equation}
\gamma = f_3^{\text{tet}} n_3^{\text{tet}} + f_4^{\text{oct}} n_4^{\text{oct}} + f_5^{\text{oct}} n_5^{\text{oct}} + f_2^{\text{O}} n_2^{\text{O}} ,
\end{equation}
where $n_3^{\text{tet}}$, $n_4^{\text{oct}}$, $n_5^{\text{oct}}$, and $n_2^{\text{O}}$ represent the numbers of under-coordinated atoms per unit area at the respective surface termination and the parameters $f_3^{\text{tet}}$, $f_4^{\text{oct}}$, $f_5^{\text{oct}}$, and $f_2^{\text{O}}$ correspond to their energy contributions. Fitting this model to our DFT-calculated surface energies yield the parameters displayed in Table~\ref{tab:model_params}, indicating that the errors of the predicted surface energies are below \SI{10}{\percent} for all stoichiometric terminations. Our results show that under-coordinated oxygen atoms and 3-fold tetrahedrally coordinated Ga2 atoms increase the surface energy substantially, suggesting that their presence destabilizes the surfaces. Notably, 5-fold octahedrally coordinated Ga1 atoms contribute negatively to the surface energy, suggesting a stabilizing effect. These results agree with computed bond-stretching force constants reported by Mu \textit{et al.}~\cite{mu2020}, where octahedral Ga1-O bonds are significantly easier to cut compared to tetrahedral Ga2-O bonds, explaining the increased stability of surfaces without dangling bonds at tetrahedral Ga2 sites such as (100). This simple model provides a qualitative framework for understanding how the atomic coordination of surface atoms determines the stability of $\beta$-Ga$_2$O$_3$ surfaces. It could also be used to predict the relative stability of higher-order stoichiometric surface orientations, which could be explored in future work.

\begin{table}[htbp]
\centering
\caption{Energy contributions of different coordination environments of under-coordinated atoms to stoichiometric surface terminations of $\beta$-Ga$_2$O$_3$.}
\label{tab:model_params}
\begin{tabular}{lr}
\hline
Coordination & Energy (J/atom) \\
\hline
3-fold Ga2 ($f_3^{\text{tet}}$) & 12.76 \\
4-fold Ga1 ($f_4^{\text{oct}}$) & 5.45 \\
5-fold Ga1 ($f_5^{\text{oct}}$) & -7.46 \\
2-fold O ($f_2^{\text{O}}$) & 18.61 \\
\hline
\end{tabular}
\end{table}

\subsection{Atomic relaxation patterns\label{sec:atomic-relaxations}}

\begin{figure} \centering
    \includegraphics[width=0.4\textwidth]{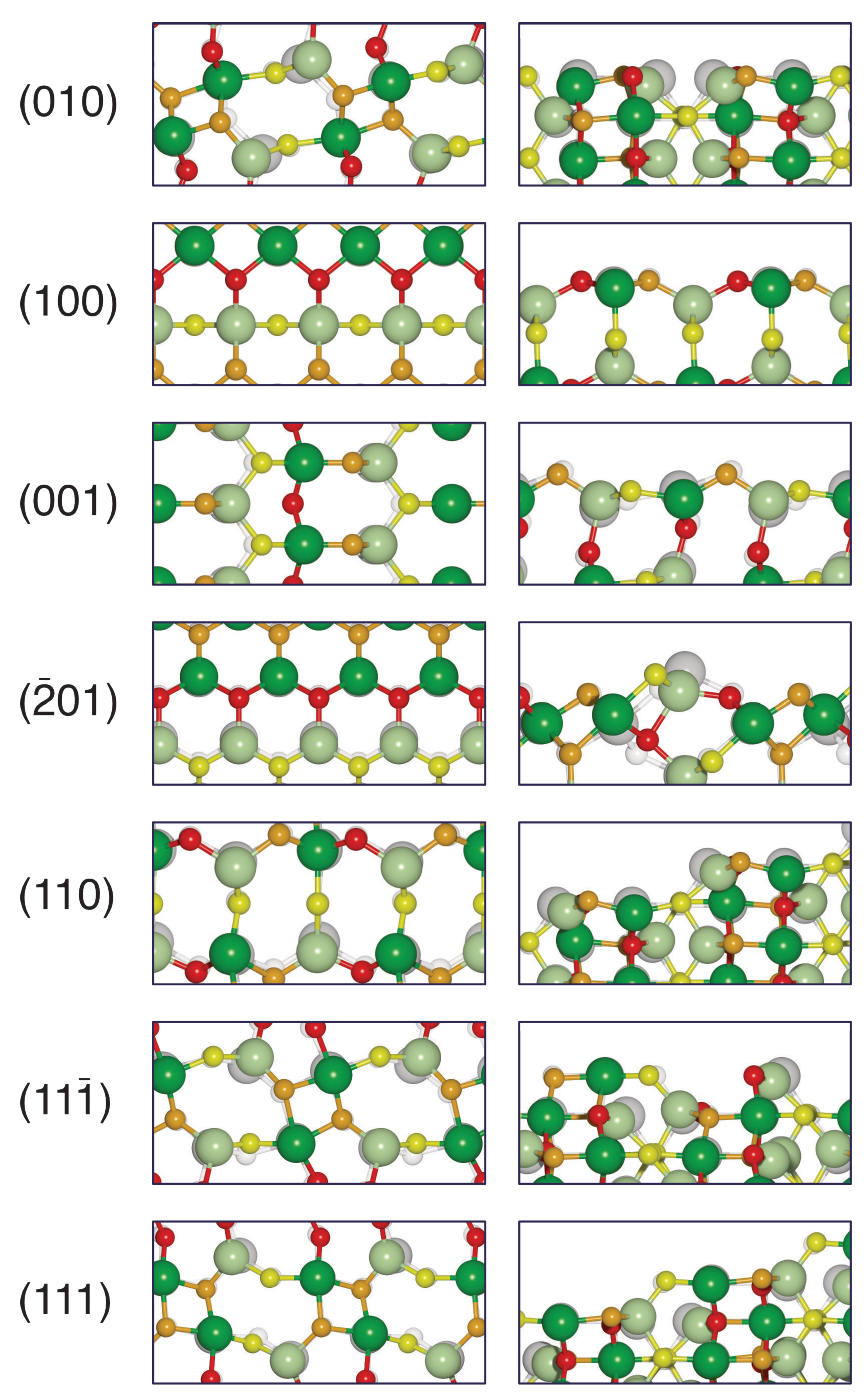}
    \caption{ Top (left) and side (right) views of the stable stoichiometric terminations of the low-index surfaces. Only the topmost surface layers are shown in the top views. The relaxed (unrelaxed) positions are shown in color (gray). \label{fig:displacements_all_strucs}}
\end{figure} 

The (100) surface exhibits minimal atomic relaxation, with displacements of all atomic species below \SI{0.1}{\angstrom}\blue{, consistent with surface x-ray diffraction data~\cite{kao2026} and with its low surface energy of \SI{0.6}{\joule\per\meter^2}}. The nearly bulk-like positions of the under-coordinated surface atoms suggest that cutting octahedral Ga1-O bonds leads to an energetically favorable configuration (see Section~\ref{sec:stoichiometric}). Figure~\ref{fig:displacements_all} presents the atomic displacements for all investigated stoichiometric terminations, decomposed into in-plane ($\Delta x$, $\Delta y$) and surface-normal ($\Delta z$) components, their visualization is shown in Fig.~\ref{fig:displacements_all_strucs}.

The ($\bar{2}$01) surface undergoes the most pronounced reconstruction among all investigated orientations, which is consistent with previously reported results~\cite{schewski2019a,mu2020}. As shown in Fig.~\ref{fig:all_terminations_individual}, the under-coordinated tetrahedral Ga2 atoms at the surface rearrange to form edge-sharing tetrahedral units. This reconstruction is accompanied by substantial subsurface relaxation, with oxygen atoms from the second layer moving upward by as much as \SI{0.5}{\angstrom} toward the surface (see Fig.~\ref{fig:all_terminations}). The edge-sharing configuration reduces the number of dangling bonds while maintaining the tetrahedral coordination preference of these Ga2 sites, explaining why this surface achieves a relatively low surface energy of \SI{0.96}{\joule\per\meter^2} despite initially presenting unfavorable under-coordinated Ga2 surface atoms.

Surfaces with higher energies show significant but localized relaxation patterns. The (010) surface displays pronounced inward relaxation of surface oxygen atoms, with $\Delta z$ displacements reaching \SI{-0.25}{\angstrom}. This inward movement reduces the exposure of under-coordinated oxygen atoms, partially compensating for the high density of dangling bonds at this termination. A similar rumpling behavior is observed for the (110), (111), and (11$\bar{1}$) surfaces, where surface cations and anions exhibit alternating inward-outward relaxation patterns to minimize electrostatic repulsion~\cite{sterrer2013}.

\subsection{Thermodynamic stability across growth conditions\label{sec:non-stoichiometric}}

\begin{figure} \centering
\includegraphics[width=0.47\textwidth]{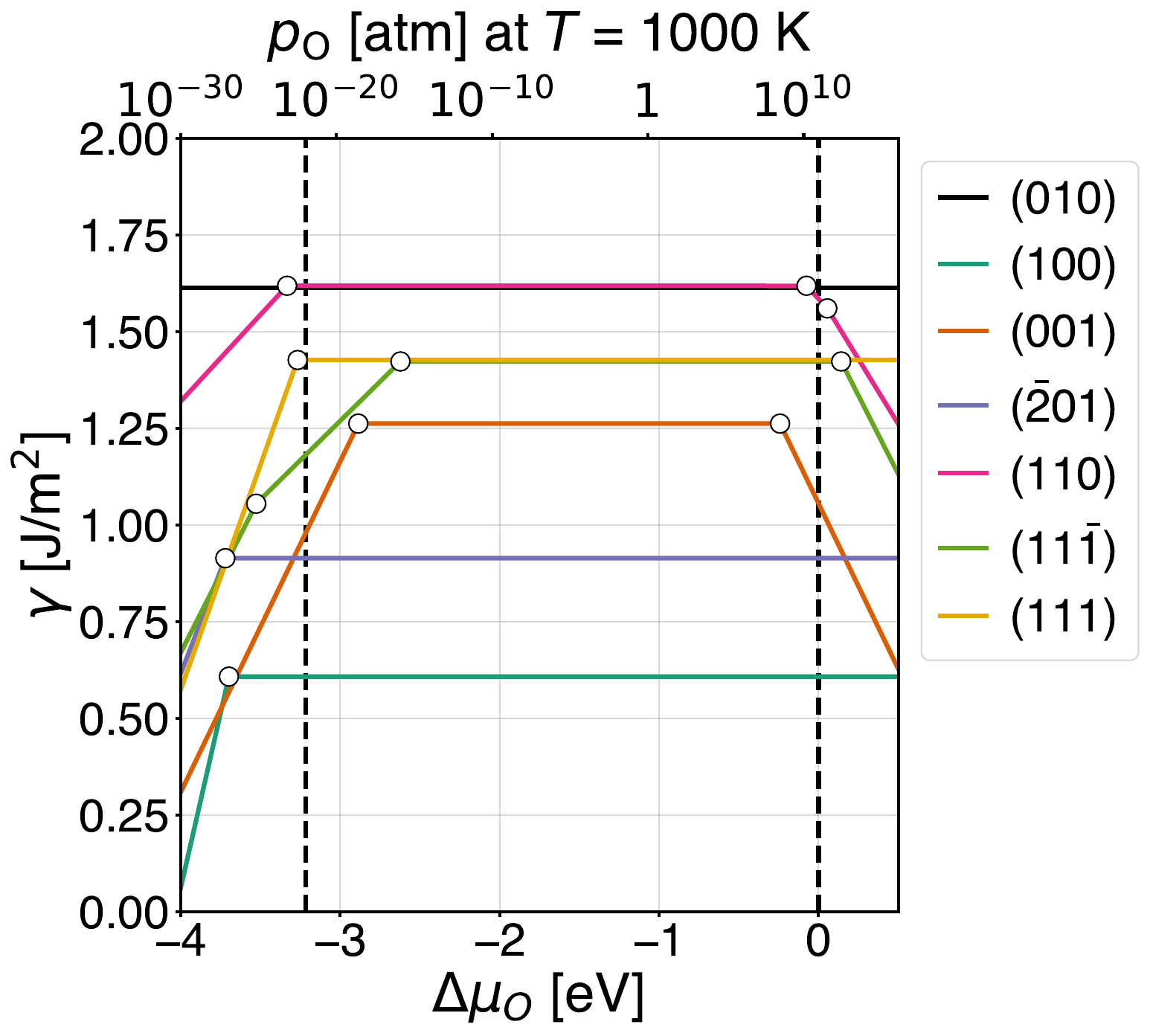}
\caption{Surface free energy diagram for stable terminations of the low-index surfaces of $\beta$-Ga$_2$O$_3$ as a function of the oxygen chemical potential $\Delta \mu_O$ obtained with PBEsol. White circles indicate transitions from one termination to another. Dashed vertical lines represent the O-poor and O-rich limits. In the top x axis, the chemical potential is transformed into a partial pressure range at a fixed temperature of \SI{1000}{\kelvin}.  \label{fig:all_terminations}}
\end{figure} 

\begin{figure}[h] \centering
    \includegraphics[width=0.4\textwidth]{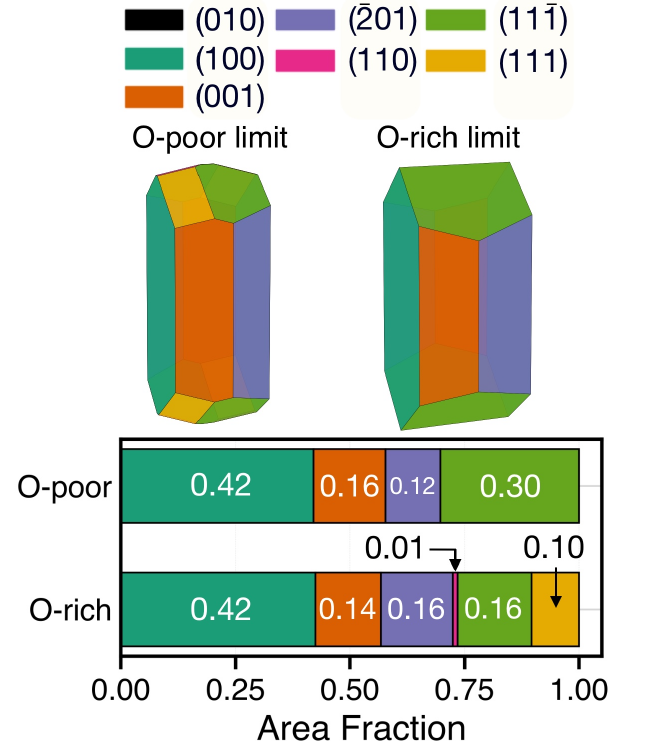}
    \caption{Predicted equilibrium shapes of $\beta$-Ga$_2$O$_3$ obtained by the Wulff construction~\cite{wulff1901,rahm2020} in the O-poor (left) and O-rich (right) limits obtained with PBEsol. The area fractions contributing to the Wulff shapes are shown in the bottom panel. \label{fig:wulffshape}}
\end{figure} 

While stoichiometric terminations provide valuable insights, real crystal growth is performed at certain oxygen partial pressures and temperatures, where non-stoichiometric terminations can become energetically favorable. We study all possible bulk-truncated terminations for all surface directions. These can be obtained by shifting the termination plane between the two inequivalent stoichiometric terminations for each surface direction~\cite{sun2013}, resulting in different O-rich and Ga-rich terminations. Figure~\ref{fig:all_terminations} presents the surface free energy diagram for all investigated surfaces as a function of the oxygen chemical potential, ranging from highly reducing (O-poor/Ga-rich) to oxidizing (O-rich/Ga-poor) conditions. The diagrams for the different surface directions, including side views of the stable terminations, are shown in Fig.~\ref{fig:all_terminations_individual}. As mentioned earlier, the (010) surface is a special case with only one stoichiometric termination and thus no other bulk-truncated terminations.

Our analysis shows that stoichiometric terminations are thermodynamically stable over a wide range of chemical potentials. For all surface orientations, the stoichiometric termination remains the lowest-energy configuration from $\Delta \mu_{\text{O}} \approx \SIrange{-2.5}{-0.5}{\electronvolt}$, spanning nearly the entire range between the thermodynamical stability limits of $\beta$-Ga$_2$O$_3$. Non-stoichiometric terminations emerge only under extreme conditions. In the highly O-poor regime ($\Delta \mu_{\text{O}} < \SI{-2.5}{\electronvolt}$), all surface directions except for (010) undergo transitions to Ga-rich terminations. Even for the most stable (100) and ($\bar{2}$01) surfaces, we find a stable Ga-rich termination that resembles a Ga adlayer structure. This supports the experimentally proposed formation of a Ga adlayer during MOVPE growth on (100) substrates~\cite{chou2023}. Chou and coworkers demonstrated that maintaining step-flow growth on (100) substrates requires low O$_2$/Ga ratios, i.e., reducing conditions, thereby lowering the Ehrlich-Schoebel barrier height~\cite{chou2023}. The same growth behavior in MOVPE may thus apply for the other surface directions.

Under O-rich conditions (\blue{$\Delta \mu_{\text{O}}>\SI{-0.5}{\electronvolt}$}), achievable through oxidizing treatments such as oxygen plasma or ozone exposure, oxygen-terminated structures begin to compete for three surface directions. The (11$\bar{1}$) surface shows a transition to an O-rich termination at $\SI{0.1}{\electronvolt}$, while the (001) surface does so near $\SI{-0.2}{\electronvolt}$. The (110) surface displays two closely competing O-rich terminations appearing near $\SI{-0.1}{\electronvolt}$, though the energy differences remain within \SI{0.05}{\joule\per\meter^2}.

The relative energetic ordering of surfaces changes with chemical potential, as evident from the crossing points in Fig.~\ref{fig:all_terminations}. In the O-poor limit ($\Delta \mu_{\text{O}} = \SI{-3.2}{\electronvolt}$), the ordering is (100) $<$ ($\bar{2}$01) $<$ (001) $<$ (11$\bar{1}$) $<$ (111) $<$ (110) $\approx$ (010), while in the O-rich limit ($\Delta \mu_{\text{O}} = \SI{0.0}{\electronvolt}$), we obtain (100) $<$ ($\bar{2}$01) $<$ (001) $<$ (11$\bar{1}$) $\approx$ (111) $<$ (110) $\approx$ (010). The Wulff construction, based on the PBEsol surface energies, predicts the equilibrium crystal shape under different growth conditions. Figure~\ref{fig:wulffshape} illustrates the equilibrium shapes in the O-poor and O-rich limits. Under O-poor conditions, the (100) facets dominate, comprising 42\% of the total surface area. The (11$\bar{1}$) and (001) surfaces contribute 30\% and 16\%, respectively, while the ($\bar{2}$01) surface accounts for 12\%. Higher-energy surfaces like (010) and (110) are absent from the equilibrium morphology under these conditions. The transition to O-rich conditions induces morphological changes while maintaining the dominance of the (100) facet (42\% area fraction). Notable changes include the appearance of the (111) facet with 10\% area fraction, compensated by a reduction in the (11$\bar{1}$) contribution to 16\%. The (001) and ($\bar{2}$01) surfaces maintain similar area fractions of 14\% and 16\%, respectively.

\section{Summary and conclusions}

We have presented a comprehensive first-principles investigation of all symmetrically inequivalent low-index surfaces of $\beta$-Ga$_2$O$_3$, examining their structural properties and thermodynamic stability across experimentally relevant growth conditions. The (010), (100), (001), ($\bar{2}$01), (110), (111), and (11$\bar{1}$) orientations were systematically studied based on density-functional theory with semi-local and hybrid functionals. Our calculations reveal that the stoichiometric surface energies span a wide range from approximately \SI{0.6}{\joule\per\meter^2} for the (100) surface to \SI{1.7}{\joule\per\meter^2} for the (111) and (010) surface. The energetic ordering remains consistent across computational approaches, demonstrating that PBEsol adequately captures the relative stability of these surfaces. The inclusion of harmonic vibrational contributions introduces modifications below \SI{0.2}{\joule\per\meter^2} up to temperatures of \SI{1000}{\kelvin} but do not alter the ordering among the investigated surfaces in terms of stability.

We have developed a coordination-based model that provides physical insight into the factors governing surface stability. Our analysis reveals that under-coordinated oxygen atoms and 3-fold tetrahedrally coordinated gallium atoms substantially destabilize surfaces, while 5-fold octahedrally coordinated gallium atoms contribute to surface stabilization. This model successfully predicts surface energies within \SI{10}{\percent} accuracy for all stoichiometric terminations and could serve as a qualitatively predictive tool for higher-order surface orientations.

The thermodynamic analysis shows that stoichiometric terminations dominate over nearly the entire range of chemical potentials relevant for $\beta$-Ga$_2$O$_3$ stability. Non-stoichiometric terminations emerge only under extreme conditions: Ga-rich terminations resembling Ga adlayers form under highly reducing conditions ($\Delta \mu_{\text{O}} < \SI{-2.5}{\electronvolt}$), while O-rich terminations appear under strongly oxidizing conditions ($\Delta \mu_{\text{O}} > \SI{-0.5}{\electronvolt}$). The prediction of stable Ga adlayer formation on (100) and ($\bar{2}$01) surfaces supports recent experimental observations during MOVPE growth under reducing conditions.

The Wulff construction based on the calculated surface energies indicates that the (100) facet dominates the equilibrium crystal morphology, accounting for approximately 42\% of the total surface area, regardless of the oxygen chemical potential. The remaining surface area is distributed among the (11$\bar{1}$), (001), ($\bar{2}$01) facets. Under O-rich conditions, the (111) facet is included. High-energy surfaces such as (010) and (110) are absent from the equilibrium shape.

While our analysis is restricted to bulk-truncated terminations, a systematic study of further reconstructions with enlarged periodicities and of adsorbate-covered structures (\eg hydrogen or water) is a natural extension of this work\blue{, as we recently demonstrated for the (001) surface, where a $1\times 2$ reconstruction of paired GaO$_4$ tetrahedra was found to be stable over a wide range of growth conditions~\cite{lion2026}}. Overall, our results provide a foundation for understanding and controlling surface properties during crystal growth and device fabrication of $\beta$-Ga$_2$O$_3$. The systematic comparison across growth conditions offers guidance for selecting appropriate surface orientations and processing parameters for specific applications in power electronics and optoelectronics.

\medskip
\section{Acknowledgements}
This work was carried out in the framework of GraFOx, a Leibniz-Science Campus partially funded by the Leibniz Association. Part of the computations were performed on the HPC systems Raven and Cobra at the Max Planck Computing and Data Facility. K.L. gratefully acknowledges discussions with Matthias Scheffler and Sergey V. Levchenko on the setup and interpretation of the surface calculations in this work.

\appendix
\counterwithin{figure}{section}
\counterwithin{table}{section}

\section{Supplementary data \label{sec:appendix}}

\begin{figure*} \centering
    \includegraphics[width=0.83\textwidth]{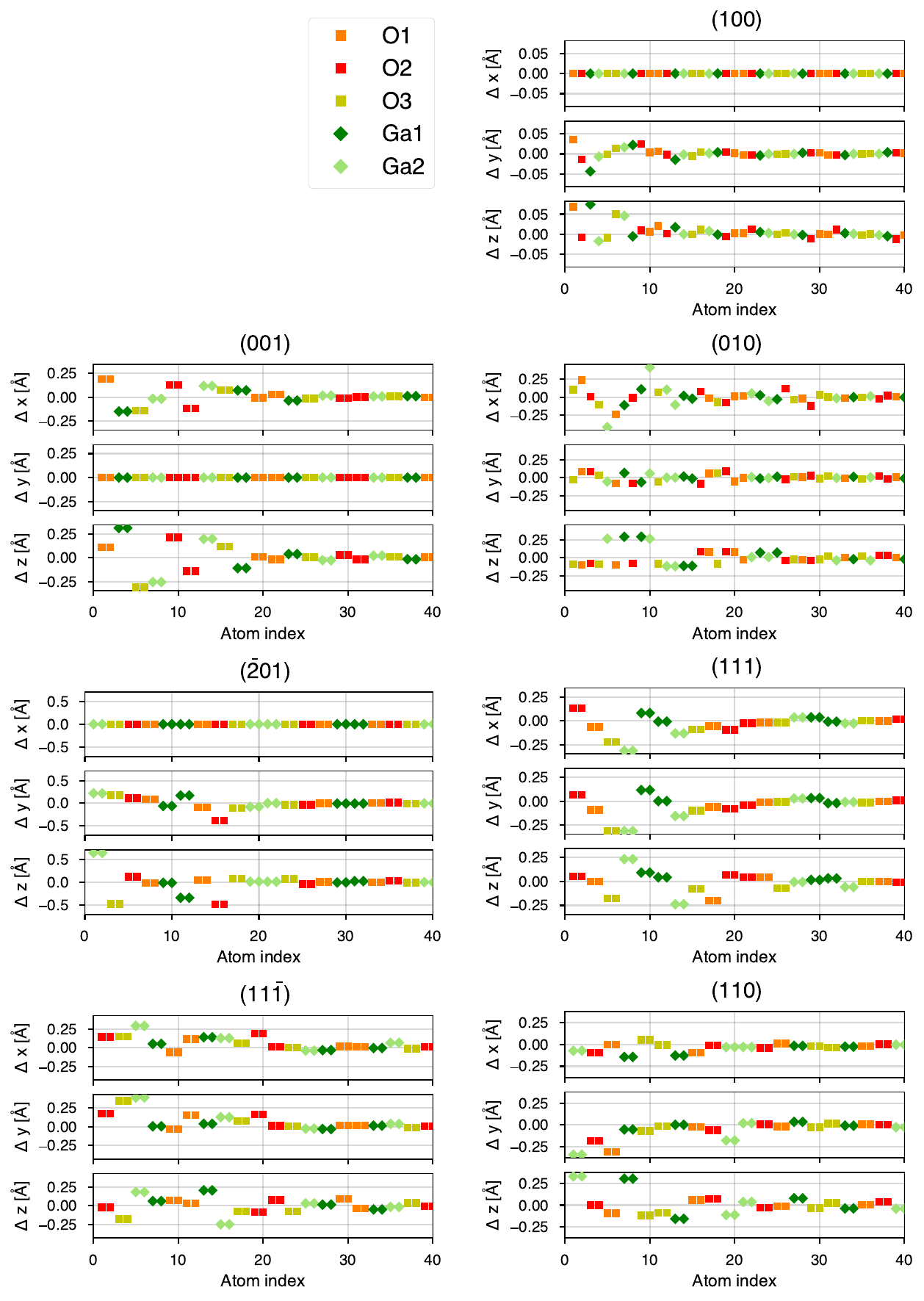}
    \caption{\blue{Atomic displacements along ($\Delta z$) and perpendicular ($\Delta x$ and $\Delta y$) the surface normal for the stable stoichiometric terminations of the low-index surfaces upon relaxation. Shown are the displacements of the first 40 atoms, starting from the top of the slab structure. \label{fig:displacements_all}}}
\end{figure*}

\begin{figure*} \centering
\includegraphics[width=0.7\textwidth]{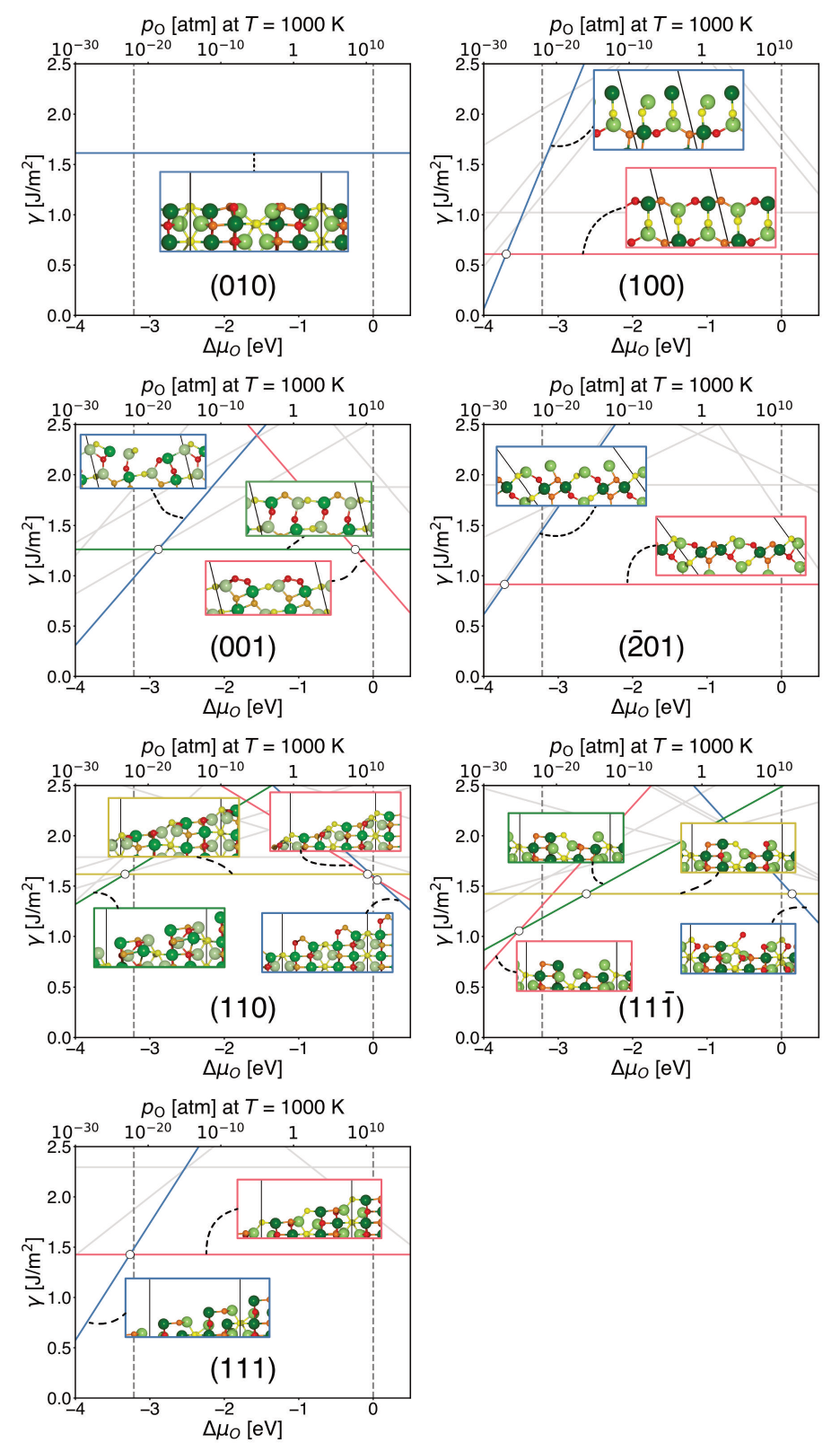}
\caption{Individual surface free energy diagrams for the low-index surfaces of $\beta$-Ga$_2$O$_3$ obtained with PBEsol. Colored lines represent terminations that have a region of stability over the entire chemical-potential range, gray lines indicate unstable terminations. Open circles reflect transitions from one stable termination to another. Side views of these stable terminations are shown as insets. The dashed vertical lines represent the O-poor and O-rich boundaries, respectively. On the top x axes, the chemical potential is converted into a partial pressure range at a fixed temperature of \SI{1000}{\kelvin}. \label{fig:all_terminations_individual}}
\end{figure*}

\begin{figure*} \centering
    \includegraphics[width=0.85\textwidth]{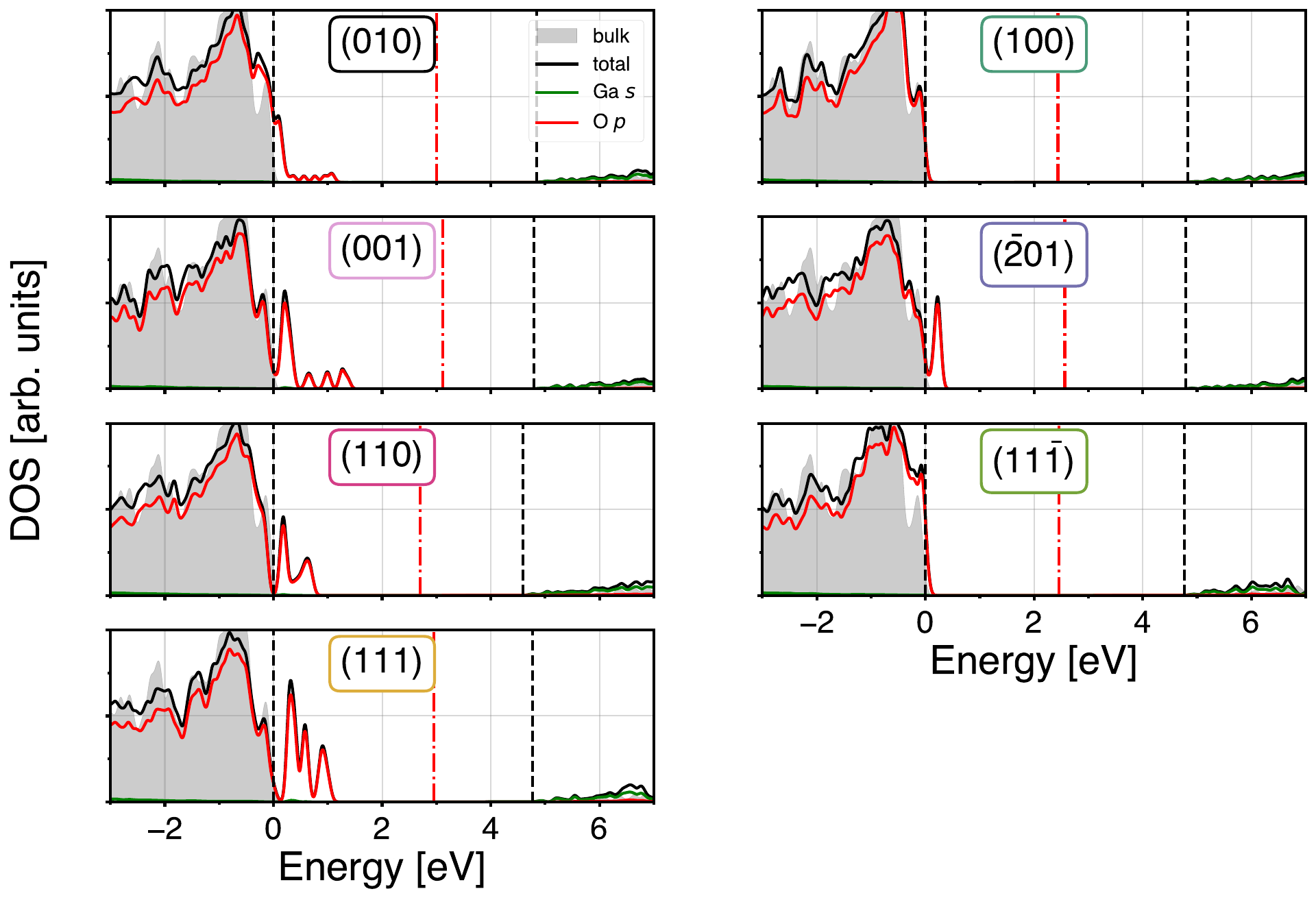}
    \caption{Density of states (DOS) and projected density of states (pDOS) of the stable stoichiometric terminations of the low-index surfaces of $\beta$-Ga$_2$O$_3$ overlayed with the bulk DOS, calculated with PBE0(0.26) at the PBEsol geometries. The energy is referenced to the VBM. The dashed black vertical lines indicate the VBM and conduction band minimum~(CBM) and the dash dotted vertical line the Fermi level. The projections onto the Ga~$s$ states are shown in green, the ones on the O~$p$ states in red. The DOS of the bulk is indicated by the shaded area. \label{fig:DOS_comparison_all_ters}}
\end{figure*}

This appendix provides supplementary data supporting the analysis presented in the main text. Figure~\ref{fig:displacements_all} presents the atomic displacements for all stoichiometric surface terminations, decomposed into in-plane and out-of-plane components, allowing for a comparison of relaxation patterns across the different surface orientations. Figure~\ref{fig:all_terminations_individual} shows the individual surface free energy diagrams for all low-index surfaces as a function of oxygen chemical potential. These diagrams provide a surface-by-surface breakdown of the thermodynamic stability analysis summarized in Fig.~\ref{fig:all_terminations} of the main text. Figure~\ref{fig:DOS_comparison_all_ters} shows the density of states (DOS) of the stable and metastable surface terminations overlayed with the bulk DOS, obtained with the PBE0(0.26) hybrid functional at the PBEsol geometries. \blue{The Fermi level lies within the band gap for all stoichiometric terminations shown in Fig.~\ref{fig:DOS_comparison_all_ters}, \ie these surfaces are insulating, consistent with their non-polar classification (see Section~\ref{sec:stoichiometric}). The states appearing above the bulk VBM are predominantly of O~$p$ character, and we assign them to the non-bonding dangling-bond states of under-coordinated oxygen atoms at the surface, which is consistent with the electron counting rule~\cite{pashley1989,wang2020}: Upon cleaving a Ga-O bond, the electrons associated with the broken bond remain primarily on the formally O$^{2-}$ oxygen atom, leaving the corresponding O~$p$-derived orbital fully occupied, while the Ga dangling bonds remain empty. Compared to the bulk, where the O~$p$ states are stabilized by bonding to three or four Ga atoms, the missing Ga-O bond and the reduced Madelung potential at the surface shift these dangling-bond states upward in energy~\cite{noguera1996}. Notably, the (010) surface exposes the highest density of under-coordinated O atoms but shows only weak in-gap features, whereas the (100) surface has a density comparable to (001) and (110) but no gap states at all. The position of the dangling-bond states is thus governed by the local environment of the under-coordinated O atoms, \ie which Ga-O bond has been cut and how many bonds remain. An analysis of the atom-projected DOS shows that pronounced states above the VBM arise only from oxygen atoms that are left two-fold coordinated, whereas oxygen atoms that retain three Ga neighbors contribute states within the bulk valence band. For the (100) and (11$\bar{1}$) terminations, only bonds of the four-fold coordinated O1 atoms to octahedral Ga1 atoms are cut, such that all surface oxygen atoms remain at least three-fold coordinated and no gap states appear. For (11$\bar{1}$) the surface atoms are additionally stabilized by the rumpling relaxation discussed in Section~\ref{sec:atomic-relaxations}. For the (001), (110), and (111) terminations, in contrast, two-fold coordinated O atoms are present, and pronounced surface states appear within the band gap, extending up to approximately \SI{1.5}{\electronvolt} above the VBM. The ($\bar{2}$01) surface is a special case. After the formation of the edge-sharing GaO$_4$ tetrahedra (Section~\ref{sec:atomic-relaxations}), the only under-coordinated atoms are the O1 atoms, which lose their bond to the tetrahedral Ga2 atoms and remain bonded to three octahedral Ga1 atoms. They give rise to a single, well-localized state about \SI{0.25}{\electronvolt} above the VBM. The (010) surface exposes two-fold coordinated O atoms of both the O2 and the O3 type, but only the O3-derived dangling bonds show up as weak in-gap features close to the VBM, consistent with the pronounced inward relaxation of its surface oxygen atoms and with the analysis of Wang \textit{et al.}~\cite{wang2020}, who found all O dangling bonds on the bare (010) surface to be filled, all Ga dangling bonds to be empty, and a surface state just above the VBM originating from the O atoms bonded to two tetrahedral Ga atoms (O3 in our notation). The presence of occupied gap states at under-coordinated oxygen atoms is in line with the coordination-based model of Section~\ref{sec:stoichiometric}, where the under-coordinated O atoms are identified as the strongest destabilizing factor.}

\clearpage

%

\end{document}